\documentclass[11pt]{article}
\usepackage{amsfonts,amscd,amssymb,amsmath,amsthm,latexsym}
\usepackage{pstricks}

\newcommand{\Q}{\mathbb Q}

\newcommand{\R}{\mathbb R}

\numberwithin{equation}{section}
\newcommand{\be}{\begin{equation}}
\newcommand{\ee}{\end{equation}}
\newcommand{\bea}{\begin{eqnarray}}
\newcommand{\eea}{\end{eqnarray}}

\date{}

\begin{document}

\title{$p$-Adic Mathematical Physics: The First 30 Years}

\author{B.~Dragovich$^{1,2}$, A.~Yu.~Khrennikov$^{3,4}$, S.~V.~Kozyrev$^5$,
\\ I.~V.~Volovich$^5$ and E. I. Zelenov$^5$ \\
$^1$Institute of Physics, University of Belgrade, 11080 Belgrade,
Serbia  \\   
$^2$Mathematical Institute, Serbian Academy of Sciences and Arts, Serbia \\
$^3$International Center for Mathematical Modeling in Physics, \\ Engineering, Economics, 
and Cognitive Science, Linnaeus University,\\ S-35195, V\"axj\"o, Sweden \\
$^4$National Research University of Information Technologies, \\ Mechanics and Optics (ITMO), 
St. Petersburg, 197101 Russia \\
$^5$Steklov Mathematical Institute of the Russian Academy of Sciences,\\ Gubkina Str. 8, Moscow,
119991 Russia}

\maketitle

\begin{abstract}
$p$-Adic mathematical physics is a branch of  modern mathematical physics  based on the application of $p$-adic mathematical methods in modeling
physical and related phenomena. It emerged in 1987  as a result of efforts to find a non-Archimedean approach to the spacetime and string dynamics at the Planck scale, but then was extended to many other areas including biology. This paper contains a brief review of main achievements in some selected topics of $p$-adic mathematical
physics and its applications, especially in the last decade. Attention is mainly paid to developments with promising  future prospects.
\end{abstract}

{\bf Key words:}{$p$-adic numbers, adeles, ultrametrics, $p$-adic mathematical physics, $p$-adic wavelets, complex systems, hierarchy, $p$-adic string theory,
quantum theory, gravity, cosmology,  stochastic processes, biological systems, proteins, genetic code, cognitive science.}







\section{{Introduction}}

{\em $p$-Adic mathematical physics} (PAMP) is a branch of  modern mathematical physics  based on the application of $p$-adic mathematical methods in modeling
physical phenomena. It is mainly related to the phenomena: $(i)$ at the very short spacetime scales and $(ii)$ of the very complex systems with hierarchical structure.
PAMP emerged in 1987 as a result of efforts to find a non-Archimedean approach to the spacetime and string dynamics at the Planck scale, see paper \cite{Volovich}, review \cite{BrFr} and book \cite{VVZ}. PAMP is not constrained only to the physical systems but includes also  related topics in mathematics and sciences, in particular in biology.

There are many physical and mathematical motivations to investigate $p$-adic mathematical physics. Below are some primary motivations.
\begin{itemize}
\item There is well-known generic quantum gravity restriction to measure distances at and beyond the Planck scale, because of the relation
\begin{equation}
\Delta x \geq \ell_P = \sqrt{\frac{\hbar G}{c^3}} \sim 10^{-33} \text{cm} \,,    \label{int.1}
\end{equation}
where  $\Delta x$ is the uncertainty measuring a distance $x$ and $\ell_P$ is the Planck length. This is a situation when the uncertainty (error) is not smaller than the value of measuring quantity. In fact, this means a breakdown of the Archimedean axiom and  application of  real numbers at the Planck scale. Namely, relation \eqref{int.1} is derived under the assumption that the whole spacetime in quantum theory and gravity can be described using only  real numbers and Archimedean geometry. However, to have better insight into spacetime structure at the Planck scale it is rather natural to use also $p$-adic numbers and adelic approach.

\item All  experimental and observational numerical data are rational numbers. Note that the field of rational numbers $\Q$ is dense in the fields of $p$-adic numbers $\Q_p$ as well as in the field of real numbers $\R .$ Recall also the local-global (Hasse-Minkowski) principle which states that, usually, when something is valid on all local fields ($\R, \Q_p$), it is also valid on the global field $\Q$. Hence, the simultaneous employment  of both   real (complex)  and $p$-adic analysis should lead to better understanding of the mathematical structure of fundamental physical laws. Moreover, it is natural to suppose that there is {\it number field invariance principle} \cite{Volovich2010}: Fundamental physical laws should be invariant under the change of the number fields.

\item Some phenomena of the complex systems have hierarchical structure. $p$-Adic numbers inherently  have ultrametric hierarchical structure and they seem to be the most adequate mathematical instrument for investigation of hierarchical systems and phenomena.

\item At the first sight one can doubt in application of $p$-adic numbers in description of physical and other phenomena, because results of measurements are not $p$-adic numbers \cite{BD-Vol}. However, recall that experimental numerical data are also not real numbers but rationals that are real and $p$-adic simultaneously. Moreover, complex numbers (in particular $i = \sqrt{-1}$) also cannot be seen in experiments, but nevertheless quantum theory is based on complex analysis. There is a part of $p$-adic analysis related to mapping $\mathbb{Q}_p \to \mathbb{R}\, (\mathbb{C})$ and it gives a possibility how one can connect $p$-adic models with quantum phenomena and hierarchical systems. Note that usually  progress in theoretical physics was related to invention of new concepts and principles, employment of adequate mathematical methods and formulation of the corresponding theoretical laws.
\end{itemize}

This paper contains a brief review of mathematical background, developments of some mathematical methods, applications in physics, biology and some other sciences.

In this review the attention is mainly paid to those topics that have got a considerable progress in the last decade, or so, and that they seem to be prospective research in the near future. For a review of the earlier developments of PAMP,  we refer readers to \cite{BrFr,VVZ} and \cite{DKKV}.

In addition to Introduction, this review contains the  sections and subsections listed below.
\begin{itemize}
\item 2. MATHEMATICAL METHODS.  \, Subsections: {\it \, 2.1 Mathematical Background; \, 2.2 Vladimirov Fractional Operator; \, 2.3 Analysis on General Ultrametric Spaces; \, 2.4 $p$-Adic Wavelets (Multidimensional $p$-adic metric; \, Multidimensinal wavelets; \, Wavelets with matrix dilations; \, Relation to spectral theory of pseudodifferential operators; \, Relation to coherent states); \, 2.5 Summation of $p$-Adic Series.}
\item  3. APPLICATIONS IN PHYSICS. \,  Subsections: {\it 3.1 $p$-Adic Strings, Fields and Quantum Theory ($p$-Adic strings; \, Quantization of the Riemann zeta-function; \, Zeta-function classical field theory; \, Zeta-function quantum field theory; \, Quantum L-functions and Langlands program; \, Zeta strings; \, Quantum mechanics); \, 3.2 Gravity and Cosmology; \, 3.3 $p$-Adic Stochastic Processes; \, 3.4 Disordered Systems and Spin Glasses; \, 3.5 Some Other Applications (Nonlinear equations and cascade models of turbulence; \, Non-Newtonian mechanics).}
\item  4. APPLICATIONS IN BIOLOGY. \, Subsections: {\it 4.1 Applications to Proteins and Genomes (Genome organization); \, 4.2 $p$-Adic Genetic Code and Bioinformation; \, 4.3 $p$-Adic Models of Cognition.}
\item  5. OTHER APPLICATIONS. \, Subsections: {\it 5.1 Data Mining; \, 5.2 Cryptography and Information Security; \, 5.3 Applications of $p$-Adic Analysis to Geology; \, 5.4 Some other Investigations.}
\item 6. CONCLUDING REMARKS.
\end{itemize}

\section{{Mathematical Methods}}

\subsection{Mathematical Background}

In this subsection we recall some usual basic notions on $p$-adic numbers, adeles and their functions.

The field $\mathbb{Q}_p$ of $p$-adic numbers (introduced by K. Hensel in 1897) is a completion of the field $\mathbb{Q}$ of rationals with respect to the $p$-adic norm: for rational number $x=p^{\gamma}a/b$, where non-zero integers $a$ and $b$ are not divisible by $p$, the norm is $|x|_p=p^{-\gamma}$. $p$-Adic numbers have natural expansions in the form of series
$$
x=\sum_{i=\gamma}^{\infty}x_i p^i,\quad x_i=0,\dots, p-1,\qquad \{x\}=\sum_{i=\gamma}^{-1}x_i p^i,
$$
where $\{x\}$ is called the fractional part of $x$ and for $x$ with the above expansion  one has $|x|_p=p^{-\gamma}$ (if $x_\gamma \ne 0$).
 $\mathbb{Z}_p = \{ x_p \in \mathbb{Q}_p : |x_p|_p \leq 1\}$ is  the ring of $p$-adic integers (unit ball with respect to the $p$-adic norm).

Bruhat-Schwartz test functions are locally constant functions with compact support. Any test function is a finite linear combination of the characteristic functions of balls. Bruhat-Schwartz generalized functions are linear continuous functionals on the space ${\mathcal{D}}(\mathbb{Q}_p)$ of test functions.

Relation of real and $p$-adic norms is given by the adelic formula of the multiplicative characters, i.e. in the simple form
$$
|x|_{\infty}\prod_p |x|_p=1,  \quad 0\neq x \in \mathbb{Q} ,
$$
where $|\cdot |_{\infty}$ is the real norm (absolute value). Adelic product of the additive characters  connects  real and all $p$-adic fractional parts of the
rational numbers, namely
$$
\chi_\infty (x) \prod_p \chi_p (x)  = \exp{(-2\pi i x)} \prod_p \exp{(2 \pi i \{ x \}_p)}     = 1 ,  \quad  x \in \mathbb{Q},
$$
where $\{ x \}_p$ denotes the fractional part of $x$.

Another relation between real and $p$-adic parameters is given by the Monna map (\cite{Monna}), subsection 2.4 below. This map and its deformations give $p$-adic parameters in complex, disordered and fractal systems, in particular in the Cantor set, spin glasses (parametrization of the Parisi matrix) and genetic code, see below.

An adele $x$ is an infinite sequence \cite{Gelfand,BD-0}
$$
 x = (x_\infty , \, x_2 , \, x_3 , \, ... , x_p , \, ...) ,
$$
where $x_\infty \in \mathbb{R}$ and $x_p \in \mathbb{Q}_p$ with the restriction that for all but a finite set $S$ of primes $p$ one has $x_p \in \mathbb{Z}_p$.
Componentwise addition and multiplication of adeles are standard  arithmetical operations on the ring of adeles $\mathbb{A}$, which can be
presented as
$$
\mathbb{A} = \bigcup_S  \mathbb{A} (S) ,   \quad   \mathbb{A} (S) = \mathbb{R} \times \prod_{p \in S} \mathbb{Q}_p \times \prod_{p \neq S} \mathbb{Z}_p .
$$
Rational numbers are naturally embedded into the space of adeles  $\mathbb{A}$. In complex-valued adelic analysis, elementary functions are of the form
$$
\varphi_S (x)  = \varphi_\infty (x_\infty) \, \prod_{p \in S} \varphi_p (x_p) \prod_{p \neq S} \Omega (|x_p|_p)  ,
$$
where $\varphi_\infty (x_\infty)$ is an infinitely differentiable function on $\mathbb{R}$ such that $|x_\infty|_\infty \, \varphi_\infty (x_\infty)  \to
 0$ as $|x_\infty|_\infty \to \infty$ for any $n = 0, 1, 2, 3, ...$, \,   $\varphi_p (x_p)$ are some locally constant functions with compact support, and
$$
\Omega (|x_p|_p)  = \begin{cases} 1 , \, \,  |x_p|_p \leq  1, \\
0 , \, \, |x_p|_p > 1 . \end{cases}
$$

$p$-Adic analysis with applications to mathematical physics and other fields attracted a lot of attention in the last thirty years. There are many books and reviews which are useful for mathematical background of $p$-adic mathematical physics, see e.g.
\cite{VVZ,Khrennikov:1994,BrFr,BD-5,Anashin,A-Khr-Sh-book,Kozyrev,Trudy_wavelets,tmf2014_1,Koch,Monna,Schikhof,Gelfand}.

\subsection{Vladimirov Fractional Operator}

The Vladimirov operator \cite{vladUMN,VVZ} of $p$-adic fractional
differentiation is defined as
$$
D^{\alpha} f(x)=\int_{\mathbb{Q}_p}
\chi(-kx)|k|_p^{\alpha}dk\widetilde{f}(k)d\mu(k),\qquad
\widetilde{f}(k)=\int_{\mathbb{Q}_p} \chi(kx)f(x)d\mu(x) .
$$
Here $f (x)$ is a complex-valued function of $p$-adic argument $x$ and
$\chi$ is the additive character: $\chi(x)=\exp \left(2\pi i
\{x\}\right)$, where $\{x\}$ is the fractional part of $x$, the integration runs with respect
to the Haar measure (with the normalization: measure of the unit ball is equal to one: $\mu(\mathbb{Z}_p)=1$).

For $\alpha>0$ the Vladimirov operator has the following integral
representation:
$$ D^{\alpha} f(x)=\frac{1}{\Gamma_p(-\alpha)}
\int_{\mathbb{Q}_p}\frac{f(x)-f(y)}{|x-y|_p^{1+\alpha}}dy,
$$
with the constant
$$\Gamma_p(-\alpha)={p^{\alpha}-1\over 1-p^{-1-\alpha}}.$$

There is a well-developed theory of the Vladimirov operator and
related constructions (spectral properties, operators on bounded
regions, analogs of elliptic and parabolic equations, a wave-type
equation, etc); see subsection  2.4    
on wavelets and \cite{VVZ,Koch,[D],Kochubei2014,[E],Alb-Kuz-Torba,Zuniga-add1,Zuniga-add2,Zuniga-add3}.

A basis of eigenvectors with compact support for the Vladimirov operator was found in \cite{Vladimirov} (the same property holds for the wavelet basis, see subsection 2.4, but the basis of \cite{Vladimirov} is different). In \cite{Doklady} it was mentioned that ultrametric pseudodifferential equations have compactly supported solutions and relation to the Anderson localization was discussed in \cite{Fyodorov}.

In \cite{vectorPDO} maps $\phi:\mathbb{Q}_p\to\mathbb{Q}_p$ were considered which are automorphisms of the tree of balls in $\mathbb{Q}_p$ (i.e. $\phi$ is a one to one map and image and inverse image of any ball is a ball). In this case the norm of the derivative $|\phi'(x)|_p$ is constant (does not depend on $x$, if the derivative exists). If the derivative does not exists, the formula below  holds but the constant does not possess interpretation as derivative. The corresponding map on functions acts as
$$
\Phi f(x)=f(\phi(x)).
$$
The following formula of pseudodifferentiation of a composite function takes place
$$
D^{\alpha}\left[\Phi f\right](x)=|\phi'(x)|_p^{\alpha}\Phi\left[D^{\alpha} f\right](x).
$$
There exists also a multidimensional generalization of this formula.

\subsection{Analysis on General Ultrametric Spaces}

The analysis of wavelets and pseudodifferential operators on
general locally compact ultrametric spaces was developed by Kozyrev in \cite{Kh-Koz2,Kh-Koz1,MathSbornik,Kozyrev}.
A pseudodifferential operator on ultrametric space $X$ is defined  as the integral operator
$$
D f(x)=\int_{X}T(\sup(x,y))(f(x)-f(y))\,d\nu(y) \, ,
$$
where $\nu$ is a Borel measure and the following integrability condition for
the integration kernel is satisfied: the kernel should be absolutely integrable at infinity
$$
\int_{d(x,y)>{R}}|T(\sup(x,y))|\,d\nu(y) \, ,
$$
where $R$ is some positive constant and $d(\cdot,\cdot)$ is the metric in $X$.

The $\sup(x,y)$ is the minimal ball in $X$ which contains both points $x$ and $y$, the
integration kernel $T$ is a function on the tree ${\cal T}(X)$ of balls in $X$. Thus $T(\sup(x,y))$ is a locally constant (for $x\ne y$) function of $x$, $y$.

Tree ${\cal T}(X)$ of balls in (locally compact) ultrametric space $X$ is defined as follows: balls with nonzero diameter or isolated points are vertices of the tree; edges connect pairs of vertices of the form (ball, maximal subball). This tree possesses a natural partial order by embedding of balls.

Ultrametric wavelet bases $\{\Psi_{Ij}\}$ on $X$ (which contain locally constant compactly supported mean zero complex valued functions) were introduced and it was
found that ultrametric wavelets are eigenvectors of ultrametric
pseudodifferential operators:
$$
T\Psi_{Ij}=\lambda_I\Psi_{Ij}.
$$
Let us note that locally compact ultrametric spaces under
consideration are completely general and do not possess any group
structure.

Here the index $I$ in the wavelet basis $\{\Psi_{Ij}\}$ enumerate balls in $X$ with non-zero diameters, a wavelet function $\Psi_{Ij}$ is a mean zero linear combination of characteristic functions of maximal subbals in $I$. The eigenvalue has the form of a series which should converge absolutely
$$
\lambda_{I}=T{(I)}
\nu(I)+\sum_{J>I} T{(J)} \nu(S(J,I)).
$$
Here $S(J,I)$ is the set obtained by elimination from ball $J$ the maximal subball containing ball $I$ (a sphere with center in $I$ corresponding to the ball $J$).

Analysis on $p$-adic infinite-dimensional spaces was developed by
Kochubei, Kaneko and Yasuda  \cite{[K],Koch,[L],[M],[N]}.
Another important class of ultrametric spaces consists of locally
compact fields of positive characteristic, see
the book of Kochubei \cite{[O]} for details.

\subsection{$p$-Adic Wavelets }
\label{sec_wavelets}

$p$-Adic wavelet theory was initiated by Kozyrev  \cite{wavelets}. For the review of $p$-adic wavelets see \cite{Trudy_wavelets}.
Important contributions in $p$-adic wavelet theory were done also by
Albeverio, Benedetto, Khrennikov, Shelkovich, Skopina, Evdokimov
\cite{wave1,Ben1,wave2,wave3,wave4,wave5,wave6,shelkovich,Alb-Kuz-Torba}.
Wavelets on locally compact abelian groups (in particular Kantor dyadic group) were considered in \cite{lang1,lang2,Farkov-1}, and wavelets on adeles were investigated in \cite{new_K1,new_K2,new_K3}.

In the $p$-adic case it is not possible to use for construction of wavelet basis translations by elements of
$\mathbb{Z}$ (these elements constitute a dense set in the unit ball). It was proposed to use instead translations by
representatives of elements of the factor group $\mathbb{Q}_p/\mathbb{Z}_p$ of the form
\begin{equation}\label{QpZp}
n=\sum_{l=k}^{-1}n_l p^l,\quad n_l=0,\dots,p-1\quad k\in\mathbb{Z},k<0.
\end{equation}

An example of $p$-adic wavelet was introduced in the form of
the product of the additive character and the characteristic function
$\Omega$ of the unit ball:
\begin{equation}\label{psi}
\psi(x)=\chi(p^{-1}x)\Omega(|x|_p),
\end{equation}
$$
\Omega(t)=\left\{
\begin{array}{rll}
1, && t\in [0,1], \\
0, && t\notin [0,1],  \\
\end{array}
\right.
\quad t\in {\mathbb R},
$$
$$
\chi(x)=\exp\left(2\pi i \{x\}\right), \,  \{x\}=\sum_{i=\gamma}^{-1} x_i p^{i}, \,  x=\sum_{i=\gamma}^{\infty} x_i p^{i}, \, x_i=0,\dots,p-1.
$$

The orthonormal basis $\{\psi_{\gamma n j}\}$ of $p$-adic wavelets
in $L^2(\mathbb{Q}_p)$ was constructed \cite{wavelets} by
translations and dilations of $\psi$:
\begin{equation}\label{wavelets}
\psi_{\gamma nj}(x)=p^{-{\gamma\over 2}} \chi(p^{\gamma-1}j
(x-p^{-\gamma}n)) \Omega(|p^{\gamma} x-n|_p),
\end{equation}
$$
\gamma\in \mathbb{Z},\quad n\in \mathbb{Q}_p/\mathbb{Z}_p,\quad j=1,\dots,p-1.
$$

Relation between real and $p$-adic wavelets is given by the $p$-adic
change of variable, or the Monna map \cite{Monna}
$$
\rho:\mathbb{Q}_p \to \mathbb{R}_+,\quad
\rho:\sum_{i=\gamma}^{\infty} x_i p^{i} \mapsto
\sum_{i=\gamma}^{\infty} x_i p^{-i-1},\quad x_i=0,\dots,p-1,\quad
\gamma \in \mathbb{Z}.
$$
This map transforms (for $p=2$) the basis $\{\psi_{\gamma n j}\}$ of $p$-adic
wavelets onto the basis of Haar wavelets on the positive half-line.

In comparison to real wavelets, $p$-adic wavelets possess some additional useful properties. The most important are relations to spectral properties of pseudodifferential operators and to orbits of groups of representations (or theory of coherent states).

\medskip

\noindent{\bf Multidimensional $p$-adic metric.} \,
The standard ultrametric in $\mathbb{Q}_p^d$ has the form
\begin{equation}\label{standard_norm}
d(x,y)=|x-y|_p={\rm max}(|x_l-y_l|_p),\quad l=1,\dots,d,
\end{equation}
$$
x=(x_1,\dots,x_d),\quad y=(y_1,\dots,y_d).
$$

A deformed metric depending on weights $q_l$ is introduced as
\begin{equation}\label{deformed_norm}
s(x,y)={\rm max}(q_l|x_l-y_l|_p),\quad l=1,\dots,d,\quad q_l> 0.
\end{equation}
The corresponding norm on $\mathbb{Q}_p^d$ we denote $\|\cdot\|$: $s(x,y)=\|x-y\|$. More general ultrametric can be obtained from (\ref{deformed_norm}) by a non-degenerate linear transformation
\begin{equation}\label{deformed_norm1}
r(x,y)=s(Ux,Uy),\qquad U\in Gl(\mathbb{Q}_p,d).
\end{equation}

A dilation with respect to ultrametric $r$ on $\mathbb{Q}_p^d$ is a linear map $\mathbb{Q}_p^d\to \mathbb{Q}_p^d$ which maps an arbitrary $r$-ball centered in zero to a maximal $r$-subball (with a center in zero) of this ball.

Let us note that for a given metric a dilation does not necessarily exist. For metric (\ref{standard_norm}) dilation is given by multiplication by $p$.

The below matrix $A$ is a dilation in $\mathbb{Q}_p^d$ with the metric $s$ given by (\ref{deformed_norm}) with the parameters satisfying $p^{-1}<q_1< q_2< \dots < q_d\le 1$.
\begin{equation}\label{A}
A=\left(\begin{array}{ccccc}
                           0 & 1 & 0 & \dots & 0\cr
                           0 & 0 & 1 & \dots & 0\cr
                           \vdots & & & & \vdots\cr
                           0 & 0 & \dots & 0 & 1\cr
                           p & 0 & \dots & 0 & 0\cr
                           \end{array}\right)=
\left(\begin{array}{cccc}1 & 0 & \dots & 0\cr
                           0 & 1 & \dots & 0\cr
                           \vdots & & & \vdots\cr
                           0 & \dots & 1 & 0 \cr
                           0 & \dots & 0 & p \cr
                           \end{array}\right)
                           \left(\begin{array}{ccccc}
                           0 & 1 & 0 & \dots & 0\cr
                           0 & 0 & 1 & \dots & 0\cr
                           \vdots & & & & \vdots\cr
                           0 & 0 & \dots & 0 & 1\cr
                           1 & 0 & \dots & 0 & 0\cr
                           \end{array}\right).
\end{equation}

\medskip

\noindent{\bf Multidimensional wavelets.} \, The set of functions $\{\psi_{k; j n}\}$ defined by (\ref{basis_1}), (\ref{basis_01}), (\ref{basis_2}) is an orthonormal basis in $L^2(\mathbb{Q}_p^d)$ (multidimensional $p$-adic wavelet basis \cite{framesdimd})

\begin{equation}\label{basis_0}
\psi_k(x)=\chi(p^{-1}k\cdot x)\Omega(|x|_p),\qquad x\in \mathbb{Q}^d_p,\qquad k\cdot x=\sum_{l=1}^d k_l x_l,
\end{equation}
the norm $|\cdot|_p$ is given by (\ref{standard_norm}),
\begin{equation}
\label{basis_1}
\psi_{k; j n}(x)=p^{-{dj\over 2}}\psi_{k}(p^{j}x-n),
\qquad x\in \mathbb{Q}_p^d,\quad
j\in\mathbb{Z},\quad n\in \mathbb{Q}_p^d/\mathbb{Z}_p^d,
\end{equation}
the index $k$ is given by the  set
\begin{equation}
\label{basis_01}
k=\left(k_1,\dots,k_d\right),\qquad k_l=0,\dots,p-1,
\end{equation}
where at least one of $k_l$ is non-zero,
\begin{equation}
\label{basis_2}
n=\left(n^{(1)},\dots,n^{(d)}\right),\qquad
n^{(l)}=\sum_{i=\beta_l}^{-1}n^{(l)}_{i}p^{i},\quad
n_i^{(l)}=0,\dots,p-1,\quad \beta_l\in\mathbb{Z}_{-}.
\end{equation}

\medskip

\noindent{\bf Wavelets with matrix dilations.} \, Let us consider the metric $s$ given by (\ref{deformed_norm}) with the parameters satisfying $p^{-1}<q_1< q_2< \dots < q_d\le 1$ and the dilation (\ref{A}). A wavelet function with matrix dilation is given by
\begin{equation}\label{wavelet1}
\Psi_k(x)=\chi\left(k\cdot A^{-1}x\right)\Omega(\| x\|),\quad k\in \mathbb{Z}_p^d/A^{*}\mathbb{Z}_p^d\backslash\{0\},
\end{equation}
and the set of representatives $k$ is given by
$$
\left(\begin{array}{c} 0\cr 1\cr 0\cr \vdots\cr 0\cr 0\cr\end{array}\right),
\left(\begin{array}{c} 0\cr 0\cr 1\cr \vdots\cr 0\cr 0\cr\end{array}\right),\dots,
\left(\begin{array}{c} 0\cr 0\cr 0\cr \vdots\cr 0\cr 1\cr\end{array}\right).
$$

Then the set of functions
\begin{equation}\label{wavelet2}
\Psi_{k;jn}(x)=p^{-{j\over 2}}\Psi_{k}(A^{j}x-n),\qquad j\in \mathbb{Z},\quad n\in \mathbb{Q}_p^d/\mathbb{Q}_p^d
\end{equation}
is an orthonormal basis in $L^2(\mathbb{Q}_p^d)$ (basis of $p$-adic wavelets with matrix dilations).

This basis was constructed in \cite{quincunxAK}. $p$-Adic quincunx basis with matrix dilations was described in \cite{KingSkopina}.

\medskip

\noindent{\bf Relation to spectral theory of pseudodifferential operators.} \,
$p$-Adic wavelets (\ref{wavelets}) are eigenvectors of the Vladimirov operator:
$$
D^{\alpha}\psi_{\gamma nj}= p^{\alpha(1-\gamma)}\psi_{\gamma nj}.
$$

This property can be generalized to much more general operators \cite{nhoper,Koz1}; including the case of general locally compact ultrametric spaces \cite{Kh-Koz2,Kh-Koz1,MathSbornik}; and multidimensional $p$-adic case \cite{Kh-Sh3,framesdimd,quincunxAK,multidim}.

Integral operators with more general kernels were considered in \cite{nhoper2,opers_bases}. For these operators wavelets were not eigenvectors, but matrices of operators in wavelet bases have non-zero matrix elements only for finite number of main diagonals.

Since $p$-adic wavelets are eigenvectors of integral operators, wavelets were used for the investigation of integral equations \cite{Koz1,Doklady,Koz-4-1,Al-Kh-Sh6,Al-Kh-Sh7,multidim,Al-Kh-Sh8,Z1,Z2,new_Z1,new_Z2,new_Z3,new_Z4,new_Z5}.

\medskip

\noindent{\bf Relation to coherent states.} \, The standard approach to the real wavelet is multiresolution analysis \cite{Meyer,Daub}. Multiresolution analysis was also developed for the $p$-adic case, where instead of translations by integers the translations by representatives (\ref{QpZp}) were used \cite{wave2,wave3,wave4,wave5,wave6,Kh-Sh3,Kh-Sh-Sk-1,Kh-Sh4,Kh-Sh6,Al-Ev-Sk-2,new_S1}.

In the $p$-adic case the multiresolution approach was generalized to more powerful construction related to coherent states for group actions
\cite{frames,framesdimd,quincunxAK}. Let us consider a group $G$ of transformations which acts in $\mathbb{Q}_p^d$ with some metric in the following way. The group maps a ball onto a ball and acts transitively on the tree of balls (with respect to the considered metric). The example is the $p$-adic affine group with transformations
$$
G(a,b)f(x)=|a|_p^{-{1\over 2}} f\left({x-b\over a}\right),\qquad a,b
\in\mathbb{Q}_p,\quad a\ne 0.
$$

The orbit of the wavelet (\ref{psi}) with respect to this action
gives the set of products of $p$-adic wavelets from the basis
$\{\psi_{\gamma n j}\}$ and $p$-roots of one. In general \cite{frames}, the orbit of generic complex-valued locally constant
mean zero compactly supported function gives a frame of $p$-adic
wavelets.

In the multidimensional case one can consider different examples of multidimensional metrics and correspondingly different groups acting on trees of balls. This approach generates different examples of $p$-adic wavelet bases and frames \cite{framesdimd}, including bases with matrix dilations \cite{quincunxAK}. In particular in this way one can reproduce the $p$-adic quincunx basis described in \cite{KingSkopina}.


\subsection{Summation of $p$-Adic Series}

Since rational numbers may be treated with respect to ordinary
and $p$-adic absolute values, the power series $\sum a_n\, x^n$ with $a_n \in \mathbb{Q}$ can be considered as real ($x \in \mathbb{R}$) as well as $p$-adic ($x \in \mathbb{Q}_p $).  Divergent series in the real case have usually good
$p$-adic convergence, see \cite{bd1} for the anharmonic oscillator and \cite{bd4} for some examples in quantum field theory. This is one of the motivations to investigate summation of some $p$-adic series, especially when the sum is a rational number
for a rational argument. It is of a special interest when this rational summation is $p$-adic invariant, i.e. when the sum does not depend
on the primes $p$. There are now many examples of the $p$-adic invariant rational summation with $n!$ in the coefficients of the series
\cite{bd2,bd3,bd4,bd5,bd6,bd7,bd8,bd9,bd10,bd11,bd12,bd13,bd14}. Recall that
$p$-adic norm of $n!$ has the property
   \begin{align}
   |n!|_p = p^{-\frac{n -s_n}{p-1}} \to 0 \, \, \text{when} \, \, n \to \infty \,,   \label{series1}
   \end{align}
where $s_n = n_0 + n_1 + ... + n_r$ is the sum of digits in the canonical expansion of number $n$ in base $p$, i.e. $n = n_0 + n_1 p + ... + n_r p^r .$
As a simple illustrative example of the $p$-adic invariant rational summation one can  refer to
 \begin{align}
\sum_{n=1}^\infty n!\ n =  1!\ 1 + 2!\ 2 + ... + n! \ n + ... = - 1 \, .       \label{series2}
\end{align}
In the proof of \eqref{series2} one can take any of the following two properties:
\begin{align}
  (i): \, \,  n!\ n = (n+1)! - n! \,, \quad \, \qquad  (ii): \, \,  \sum_{n = 1}^{N-1} n! \, n  = -1 + N! \,.   \label{series3}
\end{align}

In the sequence of papers \cite{bd5,bd6,bd7,bd8,bd9,bd10,bd12,bd13,bd14}, main attention has been paid to the  summation of various versions of   functional series of the form
\begin{equation}
S_k^\varepsilon (x) = \sum_{n = 0}^{+\infty} \varepsilon^n n! \, P_{k} (n; x)\, x^{n} \, , \quad x \in \mathbb{Z}_p \, ,\label{series4}
\end{equation}
where $ \varepsilon = \pm 1 \,, \quad k \in \mathbb{N}_0 = \mathbb{N}\cup\{ 0 \}$ and
\begin{align} \label{series5}
 P_{k} (n;x) = C_k (n)\, x^{k} + C_{k-1} (n)\, x^{k-1} + \dots + C_1 (n)\, x + C_0(n)
\end{align}
and $C_j (n), \, 0 \leq j \leq k ,$ are some polynomials in $n$ with integer-valued coefficients depending on the parameter $\varepsilon$.
Note that series \eqref{series4} is convergent when $x \in \mathbb{Z}_p$ for every $p$ and consequently is convergent for all $x \in \mathbb{Z} .$
The main interest was to find a class of polynomials $P_{k} (n;x)$ which provide $p$-adic invariant rational summation, that is $x \in \mathbb{Z}$
implies  $S_k^\varepsilon (x) \in \mathbb{Z} .$ It has been successfully done in papers \cite{bd12,bd13,bd14}. To solve this problem of the $p$-adic invariant rational summation the following auxiliary summation was considered
\begin{equation}
\sum_{i=0}^{n-1} \varepsilon^i i! \,[i^k x^k + U_k^\varepsilon (x)]\, x^i = V_k^\varepsilon(x)  + n!\, A_{k-1}^\varepsilon (n; x) \, ,  \label{series6}
\end{equation}
where $U_k^\varepsilon (x), \, V_k^\varepsilon (x), \, A_k^\varepsilon (n; x)$ are some polynomials in $x$ of the degree $k$ and the integer coefficients. Note that polynomials $A_k^\varepsilon (n; x)$ contain all information on invariant summation of series \eqref{series6} (see \cite{bd12,bd13,bd14}), in particular  holds
\begin{equation}
U_k^\varepsilon (x) = x\, A_{k-1}^\varepsilon (1; x) - \varepsilon \, A_{k-1}^\varepsilon (0; x) \, ,  \quad V_k^\varepsilon (x) = - \varepsilon A_{k-1}^\varepsilon (0; x) . \label{series7}
\end{equation}
Some sequences (see \cite{bd12,bd13,bd14}) composed of $A_k^\varepsilon (n; \pm 1)$ and $A_k^\varepsilon (n; 0)$ are related to a few interesting integer sequences, including the
the Bell numbers, presented in the Sloane on-line encyclopedia \cite{sloane}.

It is also worth noting some other investigations related to the $p$-adic invariant rational summation. In \cite{bd2}, adelic summation of the divergent
series in the real case was introduced. Some power series everywhere convergent on $\mathbb{R}$ and $\mathbb{Q}_p$ are introduced and considered in \cite{bd3}, see also \cite{bd6}. Paper \cite{bd9}  contains a  proof that the sum of the $p$-adic series
\begin{equation}
\sum_{n=0}^\infty n! \, n^k \, x^n  \, , \quad  k \in \mathbb{N}_0 \, , \,\, x \in \mathbb{N}\setminus \{ 1 \}  \label{series8}
\end{equation}
cannot be the same rational number  in $\mathbb{Z}_p$ for every $p$. It also contains conjecture that the sum of series \eqref{series8} is
a rational number if and only if $k=x = 1$. Some differential equations, whose solutions are $p$-adic power series with $n! ,$ are considered in \cite{bd11}.


\section{{Applications in Physics}}

\subsection{Strings, Fields and Quantum Theory}

\noindent {\bf $p$-Adic strings.} \,  $p$-Adic mathematical physics practically started by construction of $p$-adic analogs of the Veneziano scattering amplitudes for scalar tachyon strings. Two kinds of the scattering amplitudes were constructed: $p$-adic-valued and complex-valued. In the case of complex-valued amplitudes only the world sheet of strings is described by $p$-adic numbers, while all other ingredients are real or complex, i.e.,
\begin{equation}
A_p (a,b)  = \int_{\Q_p} |x|_p^{a-1}\, |1-x|_p^{b-1} \, dx \, = \, \frac{1-p^{a-1}}{1-p^{-a}}\frac{1-p^{b-1}}{1-p^{-b}}\frac{1-p^{c-1}}{1-p^{-c}} \,, \quad
a+b+c=1,             \label{4.1}
\end{equation}
and the  product adelic formula  $A_\infty (a,b) \prod_p A_p (a,b)  = 1 $ holds.
Mainly, this kind of $p$-adic strings has been investigated and occurred to be very useful in mathematical physics.

Theory of $p$-adic strings was pushed forward by construction of related effective Lagrangian
\begin{equation}
\mathcal{L}_p = \frac{m^D}{g^2}\, \frac{p^2}{p-1}\Big[ -\frac{1}{2}\, \varphi \, p^{-\frac{\Box}{2 m^2}}\, \varphi + \frac{1}{p+1}\, \varphi^{p+1}  \Big], \label{4.2}
\end{equation}
where $\varphi = \varphi (x)$ is a realvalued  scalar field on the real Minkowski spacetime $R^D ,$ $m$ and $g$ are real parameters, and
$\Box$ is d'Alambert operator. Lagrangian \eqref{4.2}  reproduces four-point amplitude \eqref{4.1} and also gives possibility to calculate
any higher-point one at the tree level. This Lagrangian stimulated further investigation of $p$-adic string theory, in particular the corresponding equation of motion
\begin{equation}
p^{-\frac{\Box}{2 m^2}}\, \varphi(x) = \varphi^p (x)            \label{4.3}
\end{equation}
has been studied \cite{VlaVolY,Vlad1eq,Vlad2eq,Vlad3eq,Vlad4eq,Vlad5eq}. The prime $p$  can be replaced with any natural number $n \geq 2$ and expressions  \eqref{4.2} and \eqref{4.3} still make sense.

On $p$-adic strings, see reviews \cite{BrFr,VVZ,DKKV} and references therein.
\medskip


\noindent {\bf  Quantization of the Riemann zeta-function.} \,
Quantization of the Riemann zeta-function motivated by $p$-adic string theory was proposed by Aref'eva and Volovich in \cite{AreVolZeta}.
The Riemann zeta-function is treated as a symbol of a pseudodifferential operator and  the corresponding
classical and quantum field theories are studied.  It is shown that the
Lagrangian for the zeta-function field is equivalent to the sum of the Klein-Gordon
Lagrangians with masses defined by the zeros of the Riemann zeta-function. Quantization
of the mathematics of Fermat-Wiles and the Langlands program   is also indicated.  Possible cosmological
applications of the zeta-function field theory are  discussed.

The Riemann zeta-function is defined as \be \label{R1}
\zeta(s)=\sum_{n=1}^{\infty}\frac{1}{n^s},~~s=\sigma+i\tau,~\sigma
>1 \,,\ee
and there is an Euler adelic representation

\be \label{R2} \zeta(s)=\prod_p(1-p^{-s})^{-1} .\ee  The zeta-function
admits an analytic continuation to the whole complex plane $s$
except the point $s=1$ where it has a simple pole.

Due to the Riemann hypothesis on zeros of the zeta-function a very interesting
operator is the following  pseudodifferential one,
\be \label{I2}
\zeta(\frac{1}{2}+i\Box)\,. \ee
We call this operator the {\it quantum zeta-function}.
Here  $\Box$ is the d'Alembert operator.  We also consider quantization
of the Riemann $\xi$-function and various $L$-functions.
Euclidean version of the quantum zeta-function
is $\zeta(\frac{1}{2}+i\Delta)$ where $\Delta$ is
the Laplace operator.

Under the assumption that the Riemann hypothesis is true it was shown that the
Lagrangian for the zeta-function field is equivalent to the sum of the Klein-Gordon
Lagrangians with masses defined by the zeros of the Riemann zeta-function
at the critical line. If the Riemann hypothesis is not true then the mass spectrum
of the field theory is different.

We can quantize  not only the Riemann zeta-function but also
more general $L$-functions. Quantization
of the mathematics of
 Fermat-Wiles \cite{Wil} and the Langlands program \cite{Lan}  is indicated.

One introduces the Riemann $\xi$-function \be \label{R3}
\xi(s)=\frac{s(s-1)}{2} \, \pi^{-\frac{s}{2}}\, \Gamma\big(\frac{s}{2}\big) \, \zeta(s) . \ee

The Riemann hypothesis asserts that all nontrivial zeros $\rho$ lie
at the critical line: $\rho=\frac{1}{2}+i\tau$. There is a
conjecture that all zeros are simple.

The zeros of the $\xi$-function are the same as the nontrivial zeros
of the $\zeta$-function. It is known that
 $\xi(\frac{1}{2}+i\tau)$ is real for real $\tau$ and
 is bounded.
 Locating zeros on the critical line of the (complex) zeta
 function reduces to locating
 zeros on the real line of the real function $\xi(\frac{1}{2}+i\tau)$.

\medskip

\noindent{\bf Zeta-function classical field theory.} \,
If $F(\tau)$ is  a function of a real variable $\tau$ then we define
a pseudodifferential operator $F({\Box})$ \cite{Hor} by using the Fourier
transform
\be \label{Z2}
 F({\Box})\phi (x)=\int e^{ixk}F(k^2)\tilde{\phi}(k)dk  \,.
 \ee
Here $\Box$ is the  d'Alembert operator \be \label{Z1a}
{\Box}=-\frac{\partial^2}{\partial x_0^2}+\frac{\partial^2}{\partial
x_1^2}+...+\frac{\partial^2}{\partial x_{d-1}^2}\,, \ee $\phi(x)$ is a
function from $x\in \mathbb{R}^d$, $\tilde{\phi}(k)$ is the Fourier
transform and $k^2=k_0^2-k_1^2-...-k_{d-1}^2$. One assumes that the
integral (\ref{Z2}) converges, see \cite{VlaVolY} for a
consideration of one-dimensional $p$-adic field equations.

One can introduce a natural field theory related to the real
valued function $F(\tau)=\xi (\frac{1}{2}+i\tau)$  defined by means
of the zeta-function. We consider the following Lagrangian
\be
\label{Z1L} {\cal L} = \phi\, \xi (\frac{1}{2}+i{\Box})\, \phi \,.
\ee
The integral
\be \label{Z2xi}
 \xi (\frac{1}{2}+i{\Box})\, \phi (x)=\int e^{ixk} \,
 \xi (\frac{1}{2}+ik^2) \, \tilde{\phi}(k) dk
 \ee
 converges if $\phi(x)$ is a
 decreasing function
 since $\xi(\frac{1}{2}+i\tau)$ is bounded.

 The operator
 $\xi (\frac{1}{2}+i{\Box})$ (or $\zeta (\frac{1}{2}+i{\Box})$)
 is the first quantization of the Riemann zeta-function. Similarly one
 can define operators $\zeta (\sigma+i{\Box})$
 and $\zeta (\sigma+i{\Delta})$ where $\Delta$ is the Laplace
 operator. We will obtain the second quantization of the Riemann
 zeta-function when we quantize the field $\phi (x).$

Let \be \label{Z3} \rho_n= \frac{1}{2}+im_n^2,~~\bar{\rho}_n=
\frac{1}{2}-im_n^2,~~m_n>0,~~n=1,2,... \ee be the zeros of the zeta-function at the
critical line.

There is the following
\\
{\bf Proposition.} {\it The Lagrangian (\ref{Z1L}) is equivalent to the
following Lagrangian}
\be
\label{Z1aa}
 {\cal L}^\prime =
\sum_{\epsilon, n} \eta_{\epsilon n}\psi_{\epsilon n}(\Box +\epsilon
m_n^2)\psi_{\epsilon n}\,,
\ee
{\it where $\epsilon =\pm 1$.}

Therefore, the zeros $m_n^2$ of the Riemann zeta-function
become the masses of elementary particles in the Klein-Gordon
equation.

\medskip

\noindent{\bf Zeta-function quantum field theory.} \,
To quantize the zeta-function classical field $\phi (x)$ which
satisfies the equation in the Minkowski space \be \label{QFT1}
F(\Box)\phi (x)=0 \,,\ee where $F(\Box)=\xi (\frac{1}{2}+i\Box)$ we can
try to interpret $\phi (x)$ as an operator valued distribution in a
 space $\cal{H}$ which satisfies the  equation (\ref{QFT1}). We
suppose that there is a representation of the Poincar\'e group and a
 vacuum vector $|0\rangle$ in $\cal{H}$ though the space $\cal{H}$
 might be equipped    with indefinite metric and the Lorentz invariance
 can be violated. The Wightman function
$$
W(x-y)=\langle 0|\phi (x)\phi (y)|0\rangle
$$
is a solution of the equation \be \label{QFT2} F(\Box) W (x)=0 \,.\ee
By using Proposition one can write the formal Kallen-Lehmann
\cite{BogSchir} representation
$$
W(x)=\sum_{\epsilon n}\int e^{ixk}f_{\epsilon n} (k)
\delta(k^2+\epsilon m^2_n)dk \,.
$$
A mathematical meaning of this formal expression requires a further investigation.

Quantization of the fields $\psi_{\epsilon n}$ with the Lagrangian (\ref{Z1aa}) can be
performed straightforwardly. We will obtain ordinary scalar fields as well as ghosts and
tachyons. Remind that tachyon is present in the Veneziano amplitude. It was removed by
using supersymmetry. Now we discuss an approach  how to use a Galois
group and quantum $L$-functions instead of supersymmetry to improve the spectrum.

 \medskip

\noindent{\bf Quantum  $L$-functions and Langlands program.} \,
In this section we briefly discuss $L$-functions and their
quantization \cite{AreVolZeta}. The role of (super)symmetry group here is played by
the Galois group, see \cite{Man,Var,Vol2}.

For any character $\chi$ to modulus $q$ one defines the
corresponding Dirichlet $L$-function \cite{anal-number-theory} by
setting
$$
L(s,\chi)=\sum_{n=1}^{\infty}\frac{\chi (n)}{n^s},~~(\sigma >1) .
$$
If $\chi$ is primitive then $L(s,\chi)$ has an analytic continuation
to the whole complex plane. One introduces the function
$$
\xi (s,\chi)= \big(\frac{\pi}{k}\big)^{-(s+a)/2} \, \Gamma
\big(\frac{s}{2}+\frac{a}{2}\big)\, L(s,\chi)\,,
$$
where $k$ and $a$ are some parameters.  The zeros $\rho$ lie in the
critical strip and symmetrically distributed about the critical line
$\sigma =1/2.$ It is important to notice that unless $\chi$ is real
the zeros will not necessary be symmetric about the real line.
Therefore if we quantize the $L$-function by considering the
pseudodifferential operator
$$
L(\frac{1}{2} +i\Box, \chi)\,,
$$
then we could avoid the appearance of tachyons
and/or ghosts by choosing an appropriate character $\chi$.

The Taniyama-Shimura conjecture relates elliptic curves and modular
forms. It asserts that if $E$ is an elliptic curve over
$\mathbb{Q}$, then there exists a weight-two cusp form $f$  which can
be expressed as the Fourier series
$$
f(z)=\sum a_n e^{2\pi nz}
$$
with the coefficients $a_n$ depending on the curve $E.$  Such a series is a modular form
if and only if  its Mellin transform, i.e. the Dirichlet $L$-series
$$
L(s,f)=\sum a_n n^{-s}
$$
has a holomorphic extension to the full $s$-plane and satisfies a functional equation.
For the elliptic curve $E$ we obtain the $L$-series $L(s,E)$.
The Taniyama-Shimura
conjecture was proved by Wiles and Taylor for semistable elliptic curves and it implies
Fermat`s Last Theorem \cite{Wil}.

Let $E$ be an elliptic curve defined over $\mathbb{Q},$
and $L(s)=L_K(s,E)$
be the $L$-function of $E$ over the field $K.$  The Birch and
Swinnerton-Dyer conjecture asserts that
$$
ord_{s=1}L(s)=r\,,
$$
where $r$ is  the rank of the group $E(K)$ of points of $E$ defined over $K.$
The quantum $L$-function in this case has the form
$$
L(1+i\Box)=A (i\Box)^r+...,
$$
where the leading term is expressed in
terms of the Tate-Shafarevich group.

Wiles`s work can be viewed as establishing connections between the automorphic forms and
the representation theory of the adelic groups and the Galois representations. Therefore
it can be viewed as part of the Langlands program in {\it number theory} and the
representation theory \cite{Lan} (for a recent consideration of the  {\it geometrical}
Langlands program see \cite{WGKF}).

Let $G$ be the Galois group of a Galois extension
of  $\mathbb{Q}$ and $\alpha$ a
representation of $G$ in $\mathbb{C}^n.$  There is the Artin
$L$-function $L(s,\alpha)$
associated with $\alpha.$ Artin conjectured that $L(s,\alpha)$
is entire, when $\alpha$
is irreducible, and moreover it is  automorphic: there exists a modular form $f$ such that
$$
L(s,\alpha) = L(s,f)\,.
$$
Langlands formulated the conjecture that $L(s,\alpha)$ is the $L$-function associated to
an automorphic representation of $GL(n,\mathbb{A})$, where
$\mathbb{A}=\mathbb{R}\times\prod_p' \mathbb{Q}_p$ is the ring of adeles of $\mathbb{Q}.$
Here $\mathbb{Q}_p$ is the field of $p$-adic numbers. Let $\pi$ be an automorphic
cuspidal representation of $GL(n,\mathbb{A})$, then there is an $L$-function $L(s,\pi)$
associated to $\pi.$ Langlands conjecture (general reciprocity law) states: Let $K$ be a
finite extension of $\mathbb{Q}$ with Galois group $G$ and $\alpha$ be an irreducible
representation of $G$ in $\mathbb{C}^n.$  Then there exists an automorphic cuspidal
representation $\pi_{\alpha}$ of $GL(n,\mathbb{A}_K)$ such that
$$
L(s,\alpha) = L(s,\pi_{\alpha})\,.
$$
Quantization of the $L$-function can be performed similarly to the
quantization of the Riemann zeta-function discussed above by
considering the corresponding pseudodifferential operator $L(\sigma
+i\Box)$ with some $\sigma.$

One speculates that we can not observe
the structure of spacetime at the Planck scale but feel only its
motive \cite{Vol2}.

There is a conjecture that the zeros of $L$-functions are distributed like the
eigenvalues of large random matrices from a gaussian ensemble, see \cite{RZ}. The limit
of large matrices in gauge theory (the master field) is derived in \cite{AV3}.

\medskip

\noindent{\bf Zeta strings.} \,   Extension of  Lagrangian for a single $p$-adic string \eqref{4.2} to a Lagrangian for the whole $p$-adic sector was considered by Dragovich \cite{bran1,bran2,bran3,bran4,bran5,bran6,bran7}. A few interesting  Lagrangians $L$
were constructed in the following additive way:
\begin{equation}
 L = \sum_{n=1}^\infty C_n \, \mathcal{L}_n = m^D \sum_{n=1}^\infty \frac{C_n}{g_n^2} \frac{n^2}{n-1} \, \left[ -\frac{1}{2}\, \phi \, p^{-\frac{\Box}{2 m^2}}\, \phi + \frac{1}{p+1}\, \phi^{p+1} \right] \,,   \label{4.4}
 \end{equation}
 where concrete model for a new scalar field $\phi(x)$ depends on the choice of the coefficients $C_n$ and coupling constants $g_n .$
 For each of the following coefficients $D_n$
 \begin{equation}
 \frac{C_n}{g_n^2} \frac{n^2}{n-1} = D_n \,, \quad D_n = 1 \,, \, (-1)^{n-1} \,, \, n+1 \,, \, (-1)^{n-1} (n +1) \,, \, \mu (n) \,, \, \mu (n) (n+1)  \label{4.5}
 \end{equation}
 one obtains a particular Lagrangian \eqref{4.4} containing the Riemann zeta function and nonpolynomial  potential $V(\phi) = -  L(\Box =0) .$ In
 \eqref{4.5}, $\mu (n)$ is the M\"obius function. For example, taking $D_n = (-1)^{n-1}$ and employing formula
 \begin{equation}
 \sum_{n=1}^\infty (-1)^{n-1} \, \frac{1}{n^s} = (1- 2^{1-s}) \, \zeta(s) \,, \quad s = \sigma + i \tau \,, \, \, \sigma > 0 \,,  \label{4.6}
 \end{equation}
 which has analytic continuation on the entire  complex $s$ plane,  one obtains Lagrangian
 \begin{equation}
 L = m^D \left[ - \frac{1}{2} \phi\, \big(1- 2^{1-\frac{\Box}{2 m^2}}\big) \, \zeta\big(\frac{\Box}{2 m^2}\big) \, \phi \, + \, \phi - \frac{1}{2}
 \ln (1 + \phi)^2  \right]
 \end{equation}

 Some basic  properties of the obtained Lagrangians have been investigated.  These field theory models with the Riemann zeta function are also interesting in itself.

\medskip
$p$-Adic fractal strings \cite{LapidusHung} and Dirac zeta functions \cite{Jong} are also considered.
\medskip

\noindent{\bf Quantum mechanics.} \, There are a few  $p$-adic analogs of ordinary quantum mechanics. In Vladimirov-Volovich-Zelenov  \cite{VVZ2,VVZ3}
formulation, $p$-adic quantum mechanics is  based on the following triple  $\{L^2 (\mathbb{Q}_p), W(x), U(t)\} ,$ where $L^2 (\mathbb{Q}_p)$ is the Hilbert space
of complex-valued functions of $p$-adic space and time. $W(z)$ is a unitary representation of the Heisenberg-Weyl group in $L^2 (\mathbb{Q}_p)$ and $U(t)$
is a unitary evolution operator in $L^2 (\mathbb{Q}_p)$.

Symplectic group and Heisenberg group in $p$-adic quantum mechanics are considered in \cite{VVZ,ZS}.
An early review of the developments of $p$-adic quantum mechanics is presented in \cite{VVZ}, see
also \cite{DKKV}.

Adelic quantum mechanics was introduced by Dragovich in \cite{BD-3,BD-4}, see also review \cite{BD-5}. Adelic quantum mechanics is a unification of ordinary and $p$-adic quantum mechanics for all primes $p .$  The corresponding probability amplitudes are treated by adelic path integral. It was shown  that probability amplitude for one-dimensional  systems with quadratic Lagrangian has the same form in $p$-adic and ordinary quantum mechanics, see \cite{BD-6,BD-7,BD-8}. In the case of the adelic harmonic oscillator \cite{BD-3,BD-4}, it was noted connection between some adelic states  and the functional equation for the Riemann zeta function. Decoherence in adelic quantum mechanic is considered in \cite{EZ3}.

$p$-Adic valued quantum mechanics was also considered, e. g., see \cite{KhrenQM} and references therein.

\medskip
Various aspects of $p$-adic quantum field theory are considered in many papers, see, e.g. \cite{VVZ,Missarov1,Missarov2,Gub3}.
A $p$-adic analog to AdS/CFT is investigated in \cite{Gub, Marco, GubMarco}, where an unramified extension of the $p$-adic numbers replaces Euclidean space as the boundary and a version of the Bruhat-Tits tree replaces the bulk.


\subsection{ $p$-Adic Gravity and Cosmology.}

One can argue that at the very small (Planck) scale in cosmology and in black holes the geometry of
the spacetime should be non-Archimedean
\cite{Volovich,Volovich2010}. There should be quantum fluctuations
not only of metrics and geometry but even of the number field.
Therefore, it was suggested \cite{Volovich2010} the following {\it {number
field invariance principle}}: Fundamental physical laws should be
invariant under the change of the number field. Adelic approach to the universe was emphasized by Manin \cite{Man}.

An approach to the wave function of the universe was suggested
in \cite{ADFV}. The key idea is to take into account fluctuating number fields and present the wave function in the form of an adelic product. For this purpose a $p$-adic generalization of both classical and quantum gravitational theory was proposed. Elements of $p$-adic differential geometry are described. The action and gravitation field equations over the $p$-adic number field are investigated. $p$-Adic analogs of some known solutions to the Einstein equations are presented. It was shown that the wave function for the de Sitter
minisuperspace model can be treated in the form of an adelic product
of $p$-adic counterparts. It was argued that in quantum cosmology one should consider summation only over algebraic manifolds.

In the path
integral approach to standard quantum cosmology, the starting point
is Feynman's path integral method, i.e. the amplitude to go from one
state with intrinsic metric $h_{ij}'$ and matter configuration
$\phi'$ on an initial hypersurface $\Sigma'$ to
another state with metric $h_{ij}''$ and matter
configuration $\phi''$ on a final hypersurface
$\Sigma''$ is given by a functional integral
\begin{equation}
\langle h_{ij}'',\phi'',\Sigma''|
h_{ij}',\phi',\Sigma'\rangle_\infty =
\int {\cal D}{(g_{\mu\nu})}_\infty {\cal D}(\Phi)_\infty
\chi_\infty(-S_\infty[g_{\mu\nu},\Phi])
\label {4.1v}
\end{equation}
over all four-geometries $g_{\mu\nu}$ and matter configurations
$\Phi$, which interpolate between the initial and final
configurations. In this expression $S[g_{\mu\nu},\Phi]$ is
an Einstein-Hilbert action for the gravitational and matter
fields. This action can be calculated using metric in
the standard 3+1 decomposition
\begin{equation}
ds^2=g_{\mu\nu}dx^\mu dx^\nu=-(N^2 -N_i N^i)dt^2 + 2N_i dx^i dt
+ h_{ij} dx^i dx^j,
\label{4.2v}
\end{equation}
where $N$ and $N_i$ are the lapse and shift functions, respectively.

To perform $p$-adic and adelic generalization we first make $p$-adic
counterpart of the action using form-invariance under change
of real to the $p$-adic number fields.
Then we generalize (\ref{4.1v})
and introduce $p$-adic complex-valued cosmological amplitude
\begin{equation}
\langle h_{ij}'',\phi'',\Sigma''|
h_{ij}',\phi',\Sigma'\rangle_p =
\int{\cal D}{(g_{\mu\nu})}_p{\cal D}(\Phi)_p
\chi_p(-S_p[g_{\mu\nu},\Phi]),
\label {4.3v}
\end{equation}
where $g_{\mu\nu}(x)$ and $\Phi (x)$ are the corresponding
$p$-adic counterparts of metric  and matter fields continually connecting
their values on $\Sigma'$ and $\Sigma''$. In its general aspects
$p$-adic functional integral (\ref{4.3v}) mimics the usual Feynman
path integral.
The definite integral in the classical action is understood as
the usual difference of the indefinite one at final and initial points. The measures
${\cal D}(g_{\mu\nu})_p$ and ${\cal D}(\Phi)_p$ are related to
the real-valued Haar measure on $p$-adic spaces.

Note that in (\ref{4.1v}) and (\ref{4.3v}) one has to take also a sum
over manifolds which have $\Sigma''$ and $\Sigma'$ as their boundaries.
Since the problem of topological classification of four-manifolds is
algorithmically unsolvable it was proposed that summation should
be taken over algebraic manifolds.

Investigation of $p$-adic gravity and adelic cosmology was initiated in paper \cite{ADFV}. Adelic quantum cosmology, as application of adelic quantum mechanics to minisuperspace cosmological models, was initiated by Dragovich \cite{BD-1} and has been considered in several papers, see, e.g. \cite{BD-2,DDNV,Nis1,Nis2,Nis3,Nis4}. As a result of $p$-adic effects and adelic approach to the wave function of the universe one obtains that spacetime
exhibits discreteness at the Planck scale.

Cosmological applications to the dark eergy, motivated by non-local $p$-adic effective actions and string field theory are considered in \cite{IA, VlaVolY,Calcagni,AJKV1,AJKV2,AreVolZeta,A1}.
Inspired by Lagrangian \eqref{4.2}, $p$-adic  origin  of the dark matter and dark energy was considered in  \cite{bdrag1,bdrag2}. It was also
considered  $p$-adic inflation in \cite{Barnaby}.

Nonlocality of $p$-adic string theory and string field theory was one of the motivations to consider nonlocal  gravity models with cosmological solutions (see \cite{bdrag3,bdrag4,bdrag5,bdrag6} and references therein),
given by the modified Einstein-Hilbert action in the form
\begin{equation} \label{nonlocgrav}
  S = \int \Big( \frac{R-2\Lambda}{16\pi G}+\mathcal{H}(R)\mathcal{F}(\Box)\mathcal{G}(R)\Big)\sqrt{-g}\, d^4x,
\end{equation}
where  $\mathcal{H}(R)\mathcal{F}(\Box)\mathcal{G}(R)$ is a nonlocal term with $\mathcal{F}(\Box)= \displaystyle \sum_{n =0}^{\infty}
f_{n}\Box^{n}$ as an analytic function of the d'Alembert operator $\Box .$ $\mathcal{H}(R)$ and $\mathcal{G}(R)$ are differentiable functions of
the scalar curvature $R$ and $\Lambda$ is the cosmological constant.

A  discrete stochastic model of eternal inflation based on $p$-adic numbers is proposed in \cite{Sus}.
A connection with the late-time statistical distribution of bubble-universes in the multiverse is discussed.

Algebraic geometric models in cosmology related to the "boundaries" of spacetime: Big Bang, Mixmaster Universe, Penrose's crossovers
between aeons are introduced in \cite{ManMar}. It is  suggested to model the kinematics of Big Bang using the algebraic geometric
(or analytic) blow up of a point. Symbolic dynamics, modular curves, and Bianchi IX cosmologies are considered in \cite{ManMar2,FFM}.



\subsection{$p$-Adic Stochastic Processes}

The $p$-adic diffusion (heat) equation
$$
{\partial f(x,t)\over \partial t}+D^{\alpha}_x f(x,t)=0,
$$
where $f(x,t)$ is a real valued function,  $t$ (time) is real coordinate and $x$ is $p$-adic,
$D^{\alpha}_x$ is the Vladimirov fractional operator, was suggested in \cite{VVZ}.
Mathematical properties of the corresponding stochastic processes were studied (see also \cite{[G],[H],[I],varadarajan}).

$p$-Adic Brownian motion is a real or complex valued stochastic process of $p$-adic argument $t$, defined by the stochastic pseudodifferential equation
$$
D^{\alpha}_t f(t)=\phi(t),
$$
where $\phi$ is the white noise (delta-correlated gaussian mean zero generalized stochastic process).
This stochastic process was explored in \cite{BikVol,Bikulov1999tmf,kamizono,randomfield,time series} (in particular, ultrametric analogue of the Markov property was discussed), see also \cite{evans}.

Various classes of $p$-adic stochastic processes are investigated by Albeverio and
Karwowski \cite{karwowski}, Kochubei \cite{Koch}, Yasuda
\cite{yasuda1}, Albeverio and Belopolskaya \cite{belopolskaya}; see
\cite{Koch,[C],Zuniga-add4,Zuniga-add5} for further references.  Kaneko \cite{KanFirst} showed a relationship between Besov space and
potential space. For recent developments on $p$-adic stochastic processes see \cite{processes2017_1,processes2017_2,processes2017_3,processes2017_4}.

$p$-Adic models of statistical mechanics on the Cayley tree are studied in \cite{Mukh1,Mukh2}.

\subsection{$\mathbb{Q}_p$-Valued Random Processes}

A number of articles are devoted to $\mathbb Q_p$-valued random processes or random processes with $p$-adic parameter (see, for example,
\cite{Karw1,BikVol,Khr1,Ev,Koch}). The results discussed below have been published in  \cite{EZ1,EZ2}.

Let us consider a system with $p$-adic parameter. Models of spin glasses  (\cite{Parisi,Parisi1,ABK}) and protein dynamics
(\cite{107,ABZ0,107a})  can be considered as related examples.

For an ensemble of $p$-adic systems the following natural questions arise. What is the equilibrium state of the ensemble? In what sense one should understand  deviation of the state from equilibrium one and how fast the system converges to the equilibrium?

In order to answer the above questions, the distribution of sum of the independent and identically distributed ({\em iid }) $p$-adic-valued random variables was analyzed.

Under some natural assumptions on the support of distribution of a random variable the distribution of sum of the {\em iid } variables tends to the uniform one (Haar measure) on the support of the initial distribution.
The result is well known in the framework of random walks on compact groups (\cite{answ1}).
 The initial distribution is naturally considered as close to the equilibrium one if convex hull of its support contains zero point.
If it is the case, then the distribution of sum of the {\em iid } variables converges to the uniform one. Otherwise there exist
 limits for some subsequences only. The result is known also for more general situation in the framework of $p$-adic valued  U-statistic theory (\cite{answ2}).

 The most interesting question, namely the rate of convergence of non-equilibrium distributions to the equilibrium one, is managed to obtain only for absolutely continuous distribution with locally constant density.

Let $\{\xi_n, n=1, 2, \dots\}$ be a sequence of {\em iid } random variables with values in a disk of radius $D$.
The variables are absolutely continuous and their densities  $p_\xi$ are locally constant with parameter $\Delta$. Let $p_{S_n},\, n=1, 2, \dots$ denotes the density of probability distribution of random variable  $S_n = \xi_1 +\xi_2 +\cdots + \xi_n, \, n=1, 2, \dots$. Then the following estimation is valid:
$$
\sup_{x\in\mathbb Q_p}\left|p_{S_n}(x) - \frac{1}{D}h(0,D)(x)\right|_p\leq \left(\frac{D}{\Delta}-1\right)e^{-\lambda_\xi n}.
$$
Here real number  $\lambda_\xi$ depends only on $p_\xi$  and satisfies the inequality
$$
\lambda_\xi \geq\frac{1}{3}\left(\frac{\Delta}{D}\right)^3.
$$
 The problem considered is closely related to random walks on cyclic groups (\cite{answ1}) and on profinite (commutative or noncomutative) groups
 (\cite{answ32,answ33}).

Let us define a random variable $\xi_t$  for any  $t\in\mathbb Z_p$, i. e. the family of random variables is indexed by $p$-adic integers $\{\xi_t, t\in\mathbb Z_p\}$. The function  $X(t,\omega), \, t\in\mathbb Z_p, \, \omega\in\mathbb Z_p$ coincides with random variable $\xi_t$ for any $t\in\mathbb Z_p$, \, $X(t,\omega ) = \xi_t (\omega )$. The function $X(t,\omega )$ is said to be the $p$-adic random process.

For arbitrary $p$-adic numbers $a, b$  the symbol $[a,b]$ denotes minimal disk $\mathbb Q_p$ which contains  $a, b$. We use notation $a<c<b$ for $c\in [a,b], c\neq a, c\neq b$.  Random process has independent increments iff for any finite sets
 $\{0=t_0, t_1, \dots , t_n\}$ of $p$-adic integers,  $t_0<t_1<\dots <t_n$ random variables $X(t_0, \cdot ), \, X(t_1, \cdot )-X(t_0, \cdot ), \dots , \, X(t_n, \cdot )-X(t_{n-1}, \cdot )$ are mutually independent.

$p$-Adic random process  $W(t,\cdot ) = W_t, t\in\mathbb Z_p$ is said to be ($p$-adic) Wiener  random process if it satisfies the following properties:
\begin{itemize}
\item
$W_0=0$ almost surely;
\item
$W_t$ has independent increments;
\item
 $W_t-W_s$ are absolutely continuous with densities  $\frac{1}{|t-s|_p}h\left(0,|t-s|_p\right)$.
\end{itemize}

Let $\xi_n (\omega), \, n\in\mathbb Z_+, \, \omega\in\mathbb Z_p$ be a sequence of {\em iid }  absolutely continuous random variables with constant densities  on $\mathbb Z_p$. Then the  series
\begin{equation}
\label{WP}
W(\omega , t) = \sum_{n=0}^{\infty}\xi_n(\omega) \, \gamma_n \, e_n(t)
\end{equation}
converges  for any  $t\in\mathbb Z_p$ almost surely and defines the $p$-adic Wiener process.
 Almost surely, a sample path of $p$-adic Wiener process is 1-Lipschitz function and is nowhere differentiable.

$p$-Adic Wiener process can be considered as random variable $\{W(t), t\in\mathbb Z_p\}$ with values in the space of continuous functions $C\left(\mathbb Z_p\to\mathbb Q_p\right)$. Distributions of this random variable is said to be $p$-adic Wiener measure $\mu_W$.

Let $L_0$ denotes the set of 1-Lipschits functions  with Lipschits constants equal 1 and are equal to zero at zero point.
 $L_0$ is compact $\mathbb Z_p$-modul in the space of continuous functions.
Support of  $p$-adic Wiener measure coincides with $L_0$. $p$-Adic Wiener measure under restriction on $L_0$ coincides with Haar measure on $L_0$.

In the paper \cite{Ev} the last statement is considered as definition of  $p$-adic Wiener process. It is closely related to the theory of $\mathbb Q_p$-Gaussian random variables developed in \cite{Ev}.

\subsection{Disordered Systems and Spin Glasses}

Spin glasses (disordered magnetics) are examples of disordered systems, see \cite{MPV,Talagrand,FischerHertz,BinderYoung} for the discussion of statistical mechanics of disordered systems.
Glass transition for disordered systems is described by the replica symmetry breaking approach.
The order parameter for replica symmetry breaking is given by the Parisi matrix -- some hierarchical block matrix.
It was found that \cite{MPV} the structure of correlation functions for spin glasses are related to ultrametricity.

It was shown (Avetisov, Bikulov, Kozyrev \cite{ABK} and Parisi, Sourlas \cite{PaSu}) that the Parisi matrix possesses the $p$-adic parametrization.
Matrix elements of this matrix after some natural enumeration of the indices takes the form of some real valued function of the $p$-adic norm of
difference of indices
\begin{equation}\label{Parisimat}
Q_{ab}=q(|a-b|_p).
\end{equation}
This allows to express correlation functions of spin glasses in the state with broken replica symmetry in the form of $p$-adic integrals \cite{PaSu}.

Relation of the $p$-adic Fourier transform to replica symmetry breaking was discussed in  \cite{Carlucci,Carlucci1}.
More general replica solutions related to general locally compact ultrametric spaces were obtained
\cite{ReplicaI,ReplicaII,ReplicaIII}, in this approach the Parisi matrix takes the form
\begin{equation}\label{Kozyrevmat}
Q_{ab}=q(\sup(a,b)),
\end{equation}
where the indices are enumerated by some ultrametric space and $\sup(a,b)$ is the minimal ball containing $a$, $b$ (i.e. matrix elements are generated by a real valued function on the tree of balls in some ultrametric space).

The $p$-adic Potts model was considered in \cite{mukhamedov1,mukhamedov2,mukhamedov3,mukhamedov4,mukhamedov5,mukhamedov6,mukhamedov7}.
Ultrametricity in spin glasses can be discussed as a result of branching process in multidimensional space, see for mathematical discussion \cite{NechaevVasiliev,APN_2016,Zubarev}. Ultrametricity index in spaces of high dimension was discussed in \cite{Bradley2}.

\subsection{Some Other Applications}

We mention here only a few of the other applications of $p$-adic methods in physics. Some other applications can be found in the previous
reviews \cite{BrFr,VVZ,DKKV}.

\medskip
\noindent{\bf Nonlinear equations and cascade models of turbulence.}\quad It is easy to see that the wavelet (\ref{psi}) satisfies
$$
|\psi(x)|^2\psi(x)=\psi(x).
$$

This property can be applied to construct solutions of non-linear equations, in particular, the analogue of the non-linear Schroedinger equation
\cite{Kozyrev,Al-Kh-Sh8}.

An analog  of this trick was used to construct solutions of nonlinear ultrametric integral equation \cite{[B1]} which models Richardson energy-cascading process in developed turbulence. $p$-Adic model of developed turbulence was introduced earlier by Fischenko and Zelenov \cite{[A1]} (but for their model the mentioned above trick is not applicable).

\medskip

\noindent{\bf Non-Newtionian mechanics.} \quad
Based on the assertion that Newton`s equations of motions use real numbers while one can observe only rationals,  a ``functional'' formulation of classical mechanics was suggested \cite{B1,B2,B3,B4,B5}. The fundamental equation of the microscopic dynamics in the functional approach is not the Newton equation but the Liouville  equation for the distribution function of the single particle  or the Fokker-Planck-Kolmogorov  equation. The Newton equation in this approach appears as an approximate equation describing the dynamics of the average values of the position and momenta for not too long time intervals. Corrections to the Newton trajectories are computed. An interpretation of quantum mechanics is attempted in which both classical and quantum mechanics contain fundamental randomness. Instead of an ensemble of events one introduces an ensemble of observers.

\medskip

Review of $p$-adic dynamical systems is presented in \cite{DKKV}. An adelic approach to dynamical systems is considered in \cite{BD-dyn.sys}.

Some aspects of quantum statistical mechanical systems  are considered in \cite{Marcolli1,Marcolli2}. On KMS weights on higher rank buildings, see
\cite{Marcolli3}.

\section{{Applications in Biology}}

\subsection{Applications to Proteins and Genomes}

Approximation of the space of states of a complex system by ultrametric space was proposed in theory of spin glasses and was used in biology starting from 1980-ies \cite{RTV}. In particular it was proposed (H. Frauenfelder, see \cite{Frauenfelder4,FSW,Hatom,Stein}) to consider a hierarchy of states of protein molecules.
In \cite{BeckerKarplus} the dynamics of polypeptides was modeled and it was shown that it can be approximated by a hierarchy of transitions between the energy minima described by the so called disconnectivity graph.

Models of ultrametric random walk were investigated, in particular in \cite{OgielskyStein,HoffmannSibani,BlumenKlafter}.
These earlier works were confronted with difficulties in description of different regimes in the protein dynamics, for instance in application to the experiments on ligand-rebinding kinetics of myoglobin and spectral diffusion in deeply frozen proteins.

Application of model of ultrametric diffusion generated by the Vladimirov operator to investigation of the protein dynamics was proposed in \cite{ABK,107}.
It was shown that $p$-adic diffusion equation introduced in \cite{VVZ} offers an accurate and universal description of the
protein fluctuation dynamics on extremely large range of scales from cryogenic up to room temperatures:
\begin{equation}\label{diffusion1}
{\partial f(x,t)\over \partial t}+D^{\alpha}_x f(x,t)=0,\qquad
\alpha\sim {1\over T}.
\end{equation}
Here $t$ is the real time and the $p$-adic coordinate $x$
describes the ``tree of basins'' which corresponds to the
conformational state of the protein, $T$ is the temperature. This
equation was used to describe two drastically different types of
experiments --  on rebinding of CO to myoglobin and spectral diffusion of proteins.

Applicability of this special $p$-adic diffusion equation to protein dynamics shows that the protein energy landscape exhibits the hierarchical self-similarity (which can be represented as scaling in the kernel of the Vladimirov fractional operator).

Further results on application of $p$-adic diffusion to proteins were obtained in \cite{107a,107b,107c,ABZ0,ABZ1,ABO}. In \cite{Landscape} a general diffusion on energy landscape in the Arrhenius approximation was considered (in this approximation the dynamics on the energy landscape is approximated by transitions between the energy minima). It was shown that in this case the dynamics is described by ultrametric diffusion equation with drift of the form
\begin{equation}\label{diffusion2}
\frac{\partial}{\partial t} f(x,t) +
\int_{X} \frac{e^{-\beta E(\sup(x,y))}}{\nu(\sup(x,y))}\left[e^{\beta E(x)} f(x,t)-e^{\beta E(y)}f(y,t)\right]d\nu(y)=0,
\end{equation}
where $x$ enumerates local energy minima, ultrametric space $X$ describes the ''tree of basins'' for the energy landscape ($x\in X$ are energy minima, the distance measures the transition barriers between the minima); $\sup(x,y)$ is the minimal ball in $X$ containing $x$ and $y$; $E(x)$ is the energy of minimum $x$; $\nu(y)$ measures the entropy of vicinity of energy minimum $y$; $E(\sup(x,y))$ is the energy of the Arrhenius transition state between $x$ and $y$ and $\nu(\sup(x,y))$ is the entropy of this transition state. The multipliers $e^{\beta E(x)}$, $e^{\beta E(y)}$ are drift terms, if these terms are constant (i.e. all local minima have the same energy) the diffusion generator becomes ultrametric pseudodifferential operator. Equation (\ref{diffusion2}) generalizes (\ref{diffusion1}) for the case of general energy landscape.

In \cite{ABZ2,motors} systems of $p$-adic integral equations for models of molecular motors were discussed.
A construction of molecular machines as crumpled hierarchical polymer globules was discussed in \cite{AvetisovNechaev}. In this approach the dynamics of blocks of the molecular machine is described by a system of equations controlled by a hierarchical block matrix similar to (\ref{Parisimat}) or (\ref{Kozyrevmat}).

In \cite{Nekrasov1,Nekrasov2} an approach for description of hierarchical structure of proteins was proposed. It was shown that the domain structure of proteins is one of the levels of the described hierarchy. In \cite{fragments_potentials} this approach was generalized as a model which unifies the fragment approach to proteins (construction of protein conformations as sequences of conformations of short fragments, see \cite{Frag1,Frag2,Frag3}) and the method of statistical potentials.

A model of biological evolution based on $p$-adic diffusion equation has been proposed in \cite{107d}. A model of gene duplication when genome is considered as
a probabilistic algorithm was considered in \cite{replica_algorithm}.

\medskip

\noindent{\bf Genome organization.}\quad For the structure of chromatin (organization of DNA in living cells) the structure of crumpled globule was discussed \cite{Crumpled0}. This structure is characterized by absence of entanglement of polymer chains, which is of particular importance for functioning of the
genome. One of the tricks to avoid entanglement is the ring topology for polymers \cite{RingsReview,Crumpled2}.

Another trick is the hierarchical organization of polymer globule similar to known examples of Peano curves (space filling curves) \cite{AvetisovNechaev}. This hierarchical organization was observed experimentally \cite{Crumpled3,3DgenomeReview}. The chromatin is organized in a hierarchy of levels of folding, some of levels include ring structures, this hierarchy is related to regulation of gene expression.

\subsection{$p$-Adic Genetic Code and Bioinformation}

Another very important application of $p$-adic numbers is successful modeling of the genetic code, initiated by B. Dragovich  and A. Dragovich
\cite{dragovGC1}, A. Khrennikov \cite{KHR11}, and  A. Khrennikov and S. Kozyrev \cite{genetic_code}.

Biologically, the genetic code is a connection  between codons as building blocks of the genes and the amino acids which are basic elements of the proteins. From mathematical point of view, the genetic code is a mapping from a set of $64$ elements (codons) into a set of $21$ elements ($20$ amino acids and
one stop signal). Since there are more  codons than amino acids, there is degeneration of the genetic code, i.e. a few codons  code the same amino acid.
There is an estimation that theoretically may be more than $10^{84}$ possible maps, but in the living cells there are only dozens of genetic codes.    The problem of  modeling of the genetic code consists in finding an adequate mathematical description of the existing genetic codes.

According to the B. Dragovich approach \cite{dragovGC1,dragovGC2,dragovGC3,dragovGC4,dragovGC5,dragovGC6,dragovGC7}, when two codes code the same amino acid then they are close in the information sense which can be described by the $p$-adic distance, in particular by using together both $p =5$ and $p=2 .$ The codons are numbered by the three-digit numbers in the base $p =5 ,$ where
nucleotides are identified with digits as follows: $C = 1, \, A = 2, \, T = U = 3, \, G = 4 .$ So, there is one-to-one correspondence between codons and numbers
$c_0 + c_1\, 5 + c_2\, 5^2 \equiv c_0 c_1 c_2 ,$  where $c_i = 1, 2, 3, 4.$ The vertebrate mitochondrial code is very simple code and it is presented in
Tab. 1.
 Using $p$-adic distances between codons, it was shown that degeneration of  the vertebral mitochondrial code has
$p$-adic structure, and all other codes can be regarded as slight  modifications of this one.

It is also shown that:
\begin{itemize}
 \item between codons and amino acids can be introduced $p$-adic distance \cite{dragovGC6},
 \item   not only codons but also protein amino acids make an utrametric space \cite{dragovGC6,dragovGC7},
 \item both codons and amino acids are some $p$-adic ultrametric networks \cite{dragovGC7},
\item the genetic code connects two ultrametric networks to one larger network of $85$ elements \cite{dragovGC4}.
\end{itemize}

It is also introduced \cite{dragovGC6}   the $p$-adic modified Hamming distance suitable for applications in investigation of informational similarity between sequences which elements are codons, nucleotides or amino acids. For conjecture on possible evolution of the genetic code, see \cite{dragovGC3,dragovGC4}.

\begin{table}
\caption{ The vertebrate mitochondrial code with $p$-adic  structure \cite{dragovGC7}.  $5$-Adic distance between codons is:  $\frac{1}{25}$ inside quadruplets,
 $\frac{1}{5}$ between different quadruplets in the same column, $1$ otherwise.  Each quadruplet  can be viewed as two doublets with respect to
 the $2$-adic distance between codons equal $\frac{1}{2}.$  Every doublet codes
 one amino acid or stop signal (Ter).   \label{Tab:2} }
  \small{ {\begin{tabular}{|c|c|c|c|}
 \hline \ & \ & \ & \\
  111 \, CCC \, Pro &   211 \, ACC \, Thr  &  311 \, UCC \, Ser &  411 \, GCC \, Ala  \\
  112 \, CCA \, Pro &   212 \, ACA \, Thr  &  312 \, UCA \, Ser &  412 \, GCA \, Ala  \\
  113 \, CCU \, Pro &   213 \, ACU \, Thr  &  313 \, UCU \, Ser &  413 \, GCU \, Ala  \\
  114 \, CCG \, Pro &   214 \, ACG \, Thr  &  314 \, UCG \, Ser &  414 \, GCG \, Ala  \\
 \hline \  & \  &  \ & \ \\
  121 \, CAC \, His &   221 \, AAC \, Asn  &  321 \, UAC \, Tyr &  421 \, GAC \, Asp  \\
  122 \, CAA \, Gln &   222 \, AAA \, Lys  &  322 \, UAA \, Ter &  422 \, GAA \, Glu  \\
  123 \, CAU \, His &   223 \, AAU \, Asn  &  323 \, UAU \, Tyr &  423 \, GAU \, Asp  \\
  124 \, CAG \, Gln &   224 \, AAG \, Lys  &  324 \, UAG \, Ter &  424 \, GAG \, Glu  \\
 \hline \  & \  & \  &   \\
  131 \, CUC \, Leu &   231 \, AUC \, Ile  &  331 \, UUC \, Phe &  431 \, GUC \, Val  \\
  132 \, CUA \, Leu &   232 \, AUA \, Met  &  332 \, UUA \, Leu &  432 \, GUA \, Val  \\
  133 \, CUU \, Leu &   233 \, AUU \, Ile  &  333 \, UUU \, Phe &  433 \, GUU \, Val  \\
  134 \, CUG \, Leu &   234 \, AUG \, Met  &  334 \, UUG \, Leu &  434 \, GUG \, Val  \\
 \hline \ & \   & \  &   \\
  141 \, CGC \, Arg &   241 \, AGC \, Ser  &  341 \, UGC \, Cys &  441 \, GGC \, Gly  \\
  142 \, CGA \, Arg &   242 \, AGA \, Ter  &  342 \, UGA \, Trp &  442 \, GGA \, Gly  \\
  143 \, CGU \, Arg &   243 \, AGU \, Ser  &  343 \, UGU \, Cys &  443 \, GGU \, Gly  \\
  144 \, CGG \, Arg &   244 \, AGG \, Ter  &  344 \, UGG \, Trp &  444 \, GGG \, Gly  \\
\hline
\end{tabular}}{}} 
\end{table}

\medskip

The following $2$-dimensional $2$-adic model of the genetic (amino acid) code was proposed in \cite{genetic_code}. It was shown that, if we consider the following $2$-dimensional parametrization of the nucleotides
$$
\begin{array}{|c|c|}\hline
A & G \cr\hline U & C \cr\hline
\end{array}=\begin{array}{|c|c|}\hline 00 & 01
\cr\hline 10 & 11 \cr\hline
\end{array}
$$
this gives a natural $2$-dimensional parametrization of the set of codons (the $2$-adic plane). With this enumeration the degeneracy of
the amino acid code is equivalent to local constancy of the map on the $2$-adic plane. The following table of amino acids on
the $2$-adic plane of codons was obtained:
$$
\begin{array}{|c|c|c|c|}
\hline \begin{array}{c} {\rm Lys}  \cr \hline{\rm Asn}
\end{array}  & \begin{array}{c}
{\rm Glu}  \cr \hline{\rm Asp}
\end{array}  & \begin{array}{c}
{\rm Ter}  \cr \hline{\rm Ser}
\end{array}  & {\rm Gly}\cr
\hline\begin{array}{c} {\rm Ter}  \cr \hline{\rm Tyr}
\end{array}  & \begin{array}{c}
{\rm Gln}  \cr \hline{\rm His}
\end{array}  & \begin{array}{c} {\rm Trp}  \cr
\hline{\rm Cys}
\end{array}  & {\rm Arg}\cr
\hline \begin{array}{c} {\rm Met}  \cr \hline{\rm Ile}
\end{array}
 & {\rm Val} & {\rm Thr} & {\rm Ala}    \cr
\hline
\begin{array}{c}
{\rm Leu}  \cr \hline{\rm Phe}
\end{array}
 & {\rm Leu}  & {\rm Ser}
 & {\rm Pro} \cr \hline
\end{array}
$$

Further properties of this $2$-adic $2$-dimensional parametrization were investigated
\cite{PAM,Rumer,metric_genetic_noninvariant_pANUAA,qbio} (relation to the Rumer symmetry, to PAM and BLOSUM matrices, to the energy of binding of codons, etc.). This parametrization can be considered as an example of multidimensional hierarchy, see subsection  5.1.

\subsection{$p$-Adic Models of Cognition}

 Applications of $p$-adic and more generally ultrametric spaces to mathematical modeling of cognition and psychology have quite long history, with the pioneer papers \cite{KHR0}.
The basics of the $p$-adic model of cognition were presented in monograph \cite{KHR2}; applications to psychology, including $p$-adic modeling of unconscious mind (with applications
to  Freud's  psychoanalysis and treatment of depressions) were summarized in another monograph \cite{KHR1}.  The starting point is observation that cognition is based on the hierarchic representation
of information. Hierarchy can be typically represented by using tree-like structures. Such structures can be formalized by using ultrametric spaces.  The $p$-adic trees are
the simplest trees and  $X=\mathbb{Z}_p$ is one of the simplest ultrametric spaces. Therefore it is natural to use it as the configuration space of cognition, a
{\it  mental space}.    In this model  basic mental states are represented as points of $X$ (branches of the $p$-adic tree).   One of the advantages
of operating with  $p$-adic mental space is that it can be endowed with the algebraic structure which can be used as the basis of analysis (the $p$-adic analysis).
This analysis is used to model information processing in the mental space. This model provides the possibility to encode cognitive hierarchy in
topology. Thus the $p$-adic analysis plays the role of machinery for analyzing  hierarchic information.

Real mental states are represented by {\it finite branches} and, hence, by natural numbers. The collection of real mental states represented by the set of natural numbers
$\mathbb{N}$ is embedded in the complete mental space represented by the ring of $p$-adic integers $\mathbb{Z}_p.$ Processing of mental states by the brain is mathematically modeled
with the aid of $p$-adic dynamical systems. The split of cognition into subconsciousness and consciousness plays the basic role in this model of dynamical thinking.
Powerful $p$-adic dynamical systems work in subconsciousness and we are not able to inspect consciously their iterations. Only attractors (and more generally)
cycles of these dynamical systems (unconscious information processors) are reported to conscious and then analyzed by the latter \cite{KHR0,KHR2}.

The recent development of $p$-adic and ultrametric modeling of cognition is represented in the works of G. Iurato, A. Khrennikov, and F. Murtagh \cite{FM1,FM2,FM3,I3}.
In \cite{I1,I2,I3} there was explored the $p$-adic generalization of  theory of {\it hysteresis dynamics}: conscious states are generated as the result of integrating of unconscious memories. One of
the main mathematical consequences of our model is that trees representing unconscious and conscious mental
states have to have different structures of branching.
In \cite{I1,I2,I3} there were done two new steps: a) the use of hysteresis model in the unconscious-conscious framework; b) $p$-adic mathematical
formalization of the hierarchic hysteresis. In \cite{FM1,FM2,FM3}  the ultrametric model of cognition was used for a few important applications, e.g.,
text content analysis and search and discovery, to narrative and to thinking as well as to the mathematical representation of  Matte Blanco's bi-logic.
Finally, we point to paper of Khrennikov and Kotovich \cite{KK} about ultrametric modeling of unconscious creativity


\section{{Other Applications}}

\subsection{Data Mining}
\label{sec_datamining}

One of the most promising applications of hierarchical methods is to deep learning \cite{Hinton,Bengio}.

Another well known approach in data mining is clustering, which can be considered as a hierarchical organization of data using some metric on data, see for example \cite{Murtagh,murtFirst,Murtagh1}. Hierarchical classification of data using ''tree of life'' was extensively used in biology starting from the 18-th century \cite{Linnaeus}.

Hierarchy is a natural feature in ultrametric spaces which mathematically can be expressed as a duality between ultrametric spaces and trees of balls in these spaces \cite{Lemin,Kozyrev}. In the $p$-adic case there exists also multidimensional hierarchy which is described by the Bruhat-Tits buildings
\cite{Bruhat,Garrett}.

In computer science clustering with respect to a family of metrics results in a network of clusters (this network is not necessarily a tree)
\cite{Strehl,Bauer}.  This network of clusters is obtained by combining cluster trees with respect to different metrics, where two trees are glued
together by vertices (clusters) which coincide as sets.  This approach is referred to as multi-clustering, or multiple clustering, or ensemble clustering.
One of possible applications of this approach is to phylogenetic networks needed for description of hybridization and horizontal gene transfer in biological evolution \cite{Koonin}. For the discussion of mathematical methods used in investigation of phylogenetic networks see \cite{Huson,Dress}.

It was found that network of cluster can be considered as a simplicial complex which in the case of family of metrics on multidimensional $p$-adic spaces is directly related to the Bruhat--Tits buildings \cite{cluster,hypergraph,buildings,tmf2014_1}. The example of multiclustering, or multidimensional hierarchy on the set of three points can seen on the figure (see also review \cite{tmf2014_1}).

For some other applications of $p$-adic, hierarchical and wavelet methods in data analysis and related fields see
\cite{khok,trel,bikul,joksim,Murtagh2,Murtagh3,Murtagh4,Murtagh5,Murtagh6}.

\begin{figure}\label{fig3}
\begin{pspicture}(0,4)
   \psline[linewidth=2pt](0,1.5)(0.5,2.5)(2,1.5)
   \psline[linewidth=2pt](0.5,2.5)(2,3.5)(4,1.5)
   \psline[linecolor=red,linewidth=2pt,linestyle=dashed](2,1.5)(3.5,2.5)(4,1.5)
   \psline[linecolor=red,linewidth=2pt,linestyle=dashed](3.5,2.5)(2,3.5)(0,1.5)
   \rput(0,1){\bf A}
   \rput(2,1){\bf B}
   \rput(4,1){\bf C}
   \rput(0,2){\bf AB}
   \rput(2,4){\bf ABC}
   \rput(4,3){\bf BC}
   \rput(6,2){\bf A}
   \rput(8,2){\bf B}
   \rput(8,4){\bf C}
   \psellipse[fillcolor=lightgray](7,2)(1.5,0.7)
   \psellipse[linecolor=red,linestyle=dashed](8,3)(0.7,1.5)
\end{pspicture}
\end{figure}

$p$-Adic methods were applied to classification problems in \cite{Bradley3} and to computer vision problems in \cite{Bradley1,Bradley4}.

\subsection{Cryptography and Information Security}

The possibility to use $p$-adic analysis and theory of $p$-adic dynamical systems  in cryptography was explored in a series of papers
of V. Anashin, see, e.g., \cite{A1,A2,A3}. His studies were continued in a series of joint works with A. Khrennikov and E. Yurova \cite{Anashin,MeraJNT,D1,YuRecent,SOL}.
The starting point of this approach is that the set of natural numbers $\mathbb{N}$ can be considered as a subset of the ring of $p$-adic integers
$\mathbb{Z}_p.$ Thus any data-set in the form of natural numbers can be embedded into  $\mathbb{Z}_p$ and, hence, it can be analyzed by using
the $p$-adic analysis. This is the main feature of $p$-adics explored in cryptography and information security: the possibility to treat discrete data
in the continuous framework, but continuous from the $p$-adic viewpoint. In applications we typically use $2$-adic numbers. Then
ergodic and more generally measure-preserving $p$-adic dynamical systems  are explored to encrypt information.

The main mathematical achievement
behind this approach is the possibility to formulate in practically realizable terms criteria of ergodicity and measure preserving of $p$-adic dynamical systems.
The first steps in this direction were done by V. Anashin and later new results were received  by A. Khrennikov and E. Yurova (including joint works with V. Anashin)
\cite{A3,Anashin,MeraJNT,D1,YuRecent,SOL}. Two types of such criteria are based on two special representations of maps $f: \mathbb{Z}_p \to \mathbb{Z}_p.$ One representation is through expansion of $f$
 into  the van der Put  series \cite{MeraJNT} and another representation is based on the coordinate functions associated with the $p$-adic expansion of the number $y=f(x),$ see \cite{SOL}.

Recently A. Khrennikov and E. Yurova found that $p$-adic analysis can be used as the mathematical basis for cloud computing \cite{CC0}.
Cloud computing is a type of Internet-based computing that provides shared computer processing resources and data to computers and other devices on demand.
It is a model for enabling ubiquitous, on-demand access to a shared pool of configurable
computing resources (e.g., computer networks, servers, storage, applications and services), see \cite{CC1,CC2}.

In \cite{CC0} a description of homomorphic and fully  homomorphic cryptographic primitives in the $p$-adic model was  considered. This model describes a wide class of ciphers
(including substitution ciphers, substitution ciphers streaming, keystream ciphers in the alphabet of $p$ elements), but certainly not all. Homomorphic and fully homomorphic
ciphers are used to ensure the credibility of remote computing, including cloud technology. Within considered $p$-adic model there were described all  homomorphic cryptographic
primitives  with respect to arithmetic and coordinate-wise logical operations in the ring of $p$-adic integers $\mathbb{Z}_p$. It was shown that there are no fully homomorphic
cryptographic primitives for each pair of the considered set of arithmetic and coordinate-wise logical operations on $\mathbb{Z}_p$. The problem of
 constructing a fully  homomorphic cryptographic primitives was  formulated as follows.

Consider a  homomorphic cryptographic primitives with respect to operation ``$*$"  on $\mathbb{Z}_p$. Then, it is possible to describe the complete set of operations ``$G$",
for which the cryptographic primitives are homomorphic. As a result,  a fully
homomorphic cryptographic primitives  with respect to the operations  ``$*$" and ``$G$" can be  constructed.  There was proposed the description of the set of all operations ``$G$", for which
fully  homomorphic cryptographic primitives  with respect to the operations ``$+$" and ``$G$" can be constructed  from the  homomorphic cryptographic primitive
based on the operation ``$+$".   Examples of such ``new operations'' were  presented. Of course, this work is just the first step towards creation
of $p$-adics based cloud computing. Here the expectations are very high. But the problem is extremely difficult both mathematically and methodologically.

\subsection{Application of $p$-Adic Analysis to Geology}

The cooperation between the research groups of K. Oleschko (applied geophysics and petroleum research) and A. Khrennikov ($p$-adic mathematical physics)
lead to the creation of a new promising field of research \cite{time series,O2,O3}:  $p$-adic and more generally ultrametric modeling of dynamics of flows (of, e.g., water, oil,
oil-in-water and water-in-oil droplets) in capillary networks in porous random media. The starting point of this project is the observation that
tree-like  capillary networks are very common geological structures. Fluids move through such trees of capillaries  and, hence, it is natural to try to reduce
the configuration space to these tree-like structures and the adequate mathematical model of such a configuration space is given by an ultrametric space.
The simplest tree-like structure of a capillary network can be modeled as the ring of $p$-adic integers $\mathbb{Z}_p.$

Consider the variable $x$ belonging to the field of $p$-adic numbers (or the ring of $p$-adic integers) and the  real  time variable $t.$
Here $x$ is the ``pore network coordinate'', each pathway of pore capillaries is encoded by a branch of the $p$-adic tree.
The center of this tree is selected as an arbitrary branching point of the pore  network; for a moment,
it plays the role of  the center of the coordinate, i.e., this is a purely mathematical entity; we do not assign
any geological meaning to such a center. Thus by assigning the $p$-adic number $x$ to a  system
we know in which pathway of capillaries it is located; nothing more. Hence, the $p$-adic model provides
a fuzzy description of pore networks. In particular, the size of capillaries is not included in geometry.
It can  be introduced into the model with the aid of the coefficients of the anomalous diffusion-reaction equation  playing
the role of the master equation.  From the dynamics, we get to know concentration of fluid  (oil, water or   emulsions and droplets) in capillaries.
(Recall that the dynamics are probabilistic.) However, we cannot get to know the concentration of
fluid in the precisely fixed point of the Euclidean physical space.

This modeling heavily explores theory of $p$-adic pseudodifferential equations, equations with fractional differential operators $D^\alpha$
(Vladimirov's operators), see, e.g. \cite{Koch,wavelets,nhoper,A-Khr-Sh-book,wave4,Trudy_wavelets,Kh-Koz1,Kh-Koz2,MathSbornik}.  In particular, to find solutions of $p$-adic master equations of the diffusion type one uses theory of $p$-adic wavelets
established by S. Kozyrev \cite{wavelets,nhoper}, see \cite{A-Khr-Sh-book,wave4,Trudy_wavelets} for its further development.

In spite of its mathematical beauty, the $p$-adic model does not reflect completely the complex branching structure of trees of capillaries in random porous
media. Therefore the use of general ultrametric spaces is very important for concrete geological applications. Here is used theory of ultrametric wavelets
and pseudodifferential operators which was established by  Kozyrev and  Khrennikov \cite{Kh-Koz1,Kh-Koz2,MathSbornik}. For a moment, this theory is about linear equations. However,
the real equations describing geological flows are nonlinear  both in the Euclidean \cite{R7,R9,R8} and ultramentric models \cite{O5}. There can be mentioned only  a few works
about existence of a solution of ultrametric pseudo-differential equations \cite{Kh-Koz1,Kh-Koz2,MathSbornik}. Geological studies strongly stimulate development of theory of non-linear
ultrametric pseudodifferential equations \cite{KCH}.


\subsection{Some Other Investigations}


Let us also mention some other $p$-adic subjects.

A brief review of $\mathbb{Q}_p$-valued probability is presented in \cite{DKKV}.
$p$-Adic probability logics was considered and its overview is presented in \cite{stepic}.
For a brief review of $m$-adic representation of images, see \cite{DKKV}.
For possible applications of $p$-adic analysis in econometrics, see e.g. \cite{joksim} and references therein.

See Manin's paper on numbers as functions \cite{Manin-1} and on the Painleve VI equations in $p$-adic time \cite{Manin-2}.
$p$-Adic knot invariants are discussed in \cite{Moroz}.

\section{{Concluding Remarks}}

During the past thirty years of $p$-adic mathematcal physics, a lot has been  done in invention of methods, their developments and applications.
In particular, the following methods have been introduced and well developed: the Vladimirov operator of $p$-adic fractional differentiation, various aspects of  $p$-adic wavelets, analysis on general ultrametric spaces, and summation of $p$-adic series.

Applications in physics have been mainly related to: $p$-adic strings and models with the Riemann zeta function, $p$-adic and adelic quantum mechanics, $p$-adic classical and quantum field theory,  $p$-adic gravity and cosmology, $p$-adic stochastic processes, disordered systems and spin glasses, $p$-adic dynamical systems, and some $p$-adic valued models.

Applications in biology have been mainly devoted to investigations of  the genetic code, dynamics and structure of proteins, and the hierarchical organization of polymer globules. The obtained results are in satisfactory agreement with experimental data.  A lot of work was also done towards $p$-adic modeling of the cognition.

Among other applications, there should be pointed developments in data mining, cryptography and information security, and also attempts to $p$-adic modeling
of flow dynamics in geology.

Concluding, in this paper we have presented a brief review of the recent main developments in $p$-adic mathematical physics and related topics. In particular, attention has been paid to achievements  during last decade and to those topics which are expecting  to be prospective in the near future. Some topics which are
well presented in the previous reviews \cite{BrFr,VVZ,DKKV}, are not much discussed in this article. To go into details we provide an extensive list of references (which is not the complete one).

\section*{{Acknowledgments}}

Work on this paper by one of the authors (B. Dragovich) was partially supported by Ministry of Education, Science and Technological Development of the Republic of Serbia, projects 174012
and 173052.

\end{document}